\documentclass[color]{amsart}      

\usepackage{graphicx}
\usepackage{amsmath}
\usepackage{amsfonts}
\usepackage{amssymb}
\usepackage{color}

\newcommand{\re}{{\mathbb{R}}}
\newcommand{\dd}{{\mathrm{d}}}
\newcommand{\e}{{\mathrm{e}}}

\newcommand{\V}{\mathcal{V}}
\newcommand{\D}{\mathcal{D}}

\newcommand{\Dtheta}{(\mu I+G)^{-1}}
\newcommand{\Sr}{\mathcal{S}}
\newcommand{\sgn}{\mathrm{sgn}}
\newcommand{\ipr}[2]{\langle #1, #2 \rangle}
\newcommand{\vc}[1]{{\boldsymbol{#1}}}
\newcommand{\elll}{\ell^1\text{-}\ell^2}
\newcommand{\yref}{Y}
\newcommand{\vcyref}{\vc{y}_{\text{ref}}}
\newcommand{\thetaQ}{\vc{\theta}_2^\ast}
\newcommand{\thetaS}{\vc{\theta}_{\text{sparse}}^\ast}
\newcommand{\etaS}{\vc{\eta}_{\text{sparse}}^\ast}
\newcommand{\emp}{{\bf{E}}_2}
\newcommand{\reg}{{\bf{\Omega}}}
\newcommand{\J}{{\bf{J}}}
\newtheorem{prob}{Problem}

\begin{document}
\title[Sparse Command Generator for Remote Control]{Sparse Command Generator for Remote Control%
\footnote{}
}

\author[M. Nagahara]{Masaaki Nagahara}
\author[D. E. Quevedo]{Daniel E. Quevedo}
\author[J. {\O}stergaard]{Jan {\O}stergaard}
\author[T. Matsuda]{Takahiro Matsuda}
\author[K. Hayashi]{Kazunori Hayashi}
\address{M. Nagahara and K. Hayashi are with Kyoto University.
D. E. Quevedo is with The University of Newcastle.
J. {\O}stergaard is with Aalborg University.
T. Matsuda is with Osaka University.
The corresponding author is M. Nagahara (nagahara@ieee.org).
}
\thanks{This research was supported in part by a Grand-in-Aid for
Young Scientists (B) of the Ministry of Education, Culture, 
Sports, Science and Technology (MEXT) under Grant
No. 22700069, No. 22760317, and No. 21760289,
and an Australian Research Council Australian Research Fellowship
(project number DP0988601).}

\maketitle

\begin{abstract}
In this article, we consider remote-controlled systems, where the
command generator and the controlled object are connected with a
bandwidth-limited communication link. In the remote-controlled systems, efficient 
representation of control commands is one of the crucial issues because
of the bandwidth limitations of the link.
We propose a new representation method for control commands based on
compressed sensing. In the proposed method, compressed sensing
reduces the number of bits in each control signal by representing it as
a sparse vector. The compressed sensing problem is solved by an $\elll$
optimization, which can be effectively implemented with an {\em
iterative shrinkage algorithm}. 
A design example also shows the effectiveness of the proposed method. 
\end{abstract}

\section{Introduction}
Compressed sensing has recently been a focus of intensive researches
in the signal processing community.
It aims at reconstructing a signal by assuming that the original signal is sparse
\cite{CanWak08}.
The core idea used in this area is to introduce a sparsity index
in the optimization.
The sparsity index of a vector $\vc{v}$ is defined by the amount of
nonzero elements in $\vc{v}$ and is usually denoted by $\|\vc{v}\|_0$,
called the ``$\ell^0$ norm.''
The compressed sensing decoding problem is then formulated by
least squares with $\ell^0$-norm regularization.
The associated  optimization problem is however hard to solve, since it is a
combinatorial one. Thus, it is common to introduce a convex
relaxation by replacing the $\ell^0$ norm with the $\ell^1$ norm
\cite{CheDonSau98}. Under some assumptions, the solution of this relaxed
optimization is known to be exactly the same as that of the
$\ell^0$-norm regularization \cite{ElaBru02,CanWak08}. 
That is, by minimizing the $\ell^1$-regularized least squares, 
or by {\em $\elll$ optimization},
one can obtain a sparse solution.
Moreover, recent studies have examined fast algorithms for $\elll$ optimization
\cite{DauDefMol04,BecTeb09,ZibEla10}.

The purpose of this paper is to investigate the use of sparsity-inducing
techniques for remote control \cite{Say}, see \cite{NagQue11} for an
alternative approach. In remote-controlled systems,
control information is transmitted through bandwidth-limited channels
such as wireless channels \cite{WinHol00} or the Internet
\cite{LuoChe00}. 
There are two approaches to reduce the number of bits transmitted on a
wireless link, {\em source coding} and {\em channel coding}
approaches \cite{CovTho}. In the former, information compression techniques reduce
the number of bits to be transmitted. In the latter, efficient forward
error-correcting codes reduce redundant data (i.e., parity) in
channel-coded information. In this paper, we study the former approach
and propose a sparsity-inducing technique to produce sparse
representation  of control commands, which can reduce the number of bits
in transmitted data.

Our optimization to obtain sparse representation of control commands
is formulated as follows:
we measure the tracking error in the output trajectory
of a controlled system by its $\ell^2$ norm,
and add an $\ell^1$ penalty to achieve sparsity of transmitted vector.
This is an $\ell^1$-regularized $\ell^2$-optimization,
or shortly $\elll$-optimization,
which is effectively solved by the iterative shrinkage method
mentioned above.
The problem of command generator has been solved
when the penalty is taken solely as an
$\ell^2$ norm, the solution of which is given by a linear combination
of base functions, called control theoretic splines \cite{SunEgeMar00}.
In this work, we also present a simple method for achieving sparse control vectors
when the control commands are assumed to be in a subspace of these splines.
An example illustrates the effectiveness of our method
compared with the $\ell^2$ optimization.

\subsection*{Notation}
For a vector $\vc{v}=[v_1,\ldots,v_n]^\top\in\re^n$,
the $\ell^1$ and $\ell^2$ norms are respectively defined by
$\|\vc{v}\|_1 := \sum_{i=1}^n |v_i|$ and $\|\vc{v}\|_2 := \sqrt{\vc{v}^\top\vc{v}}$.
For a real number $x\in\re$,
\[
  \sgn(x) := \begin{cases} 1,\quad \text{if}~~x\geq 0,\\ -1, \quad \text{if}~~x<0,\end{cases},~
  (x)_+ := \max\{x,0\}.
\]
We denote the determinant of a square matrix $M$ by $\det(M)$,
and the maximum eigenvalue of a symmetric matrix $M$ by $\lambda_{\max}(M)$.
Let $L^2[0,T]$ be the set of Lebesgue square integrable functions on $[0,T]$.
For $f,g\in L^2[0,T]$, the inner product is defined by
\[
\ipr{f}{g}:=\int_0^T f(t)g(t)\dd t.
\]

\section{Command Generation Problem}
\label{sec:prob}
Let us consider the following linear SISO (Single-Input Single-Output) plant:
\begin{equation}
 P:\left\{\begin{split}
 \dot{\vc{x}}(t) &= A\vc{x}(t) + \vc{b}u(t), & \\
 y(t) &= \vc{c}^\top\vc{x}(t), \quad t\in[0,\infty), \quad x(0)=0,
 \end{split}\right.
 \label{eq:system}
\end{equation}
where $A\in\re^{n\times n}$, $\vc{b}\in\re^{n}$ and $\vc{c}\in\re^{n}$.
We assume that
the system $P$ is stable and
the state space realization (\ref{eq:system})
is reachable and observable.
The output reference signal is given by
data points
$\D := \{(t_1, \yref_1), (t_2, \yref_1), \ldots (t_N, \yref_N)\}$,
where $t_i$'s are time instants such that
$0<t_1<t_2<\cdots<t_N=:T$.
Our objective here is to design the control signal $u(t)$
such that the output trajectory $y(t)$ is close to the data points $\yref_1$,\ldots,$\yref_N$
at $t=t_1,\ldots, t_N$, that is, $y(t_i)\approx\yref_i$, $i=1,\ldots, N$.
To measure the difference between $\{y(t_i)\}_{i=1}^N$ and $\{Y_i\}_{i=1}^N$,
we adopt the square-error cost function
\[
 \emp(u) = \sum_{i=1}^N (y(t_i)-Y_i)^2,
\]
where we have made the dependence of $y(t_i)$ on
$u=\{u(t)\}_{t\in[0,T]}$
through the system equation
(\ref{eq:system}).

In principle, one can achieve perfect tracking,
that is, $\emp=0$, by some input signal\footnote{
The explicit form of this input is given by (\ref{eq:optimal_u2}) and (\ref{eq:theta2})
in Section \ref{sec:l2}, with $\mu=0$.}.
However, the optimal input for perfect tracking has very large gain especially
when the number $N$ is very large,
and may lead to oscillation between the sampling instants $t_1,\ldots,t_N$.
This phenomenon is known as overfitting \cite{SchSmo}.
To avoid this, one can adopt a {\em regularization} or {\em smoothing} technique.
This method is to add a regularization term $\reg(u)$ to the cost function $\emp(u)$.
We formulate our problem as follows:
\begin{prob}
Given data $\D$, find a control signal $u$ which minimizes
the regularized cost function $\J(u) = \emp(u) + \mu \reg(u)$,
where $\mu>0$ is the regularization parameter which specifies the tradeoff
between minimization $\emp(u)$ and the smoothness by $\reg(u)$.
\end{prob}

A well-known regularization is to use $L^2$ function for $\reg(u)$,
called the {\em control theoretic smoothing spline} \cite{SunEgeMar00,EgeMar}.
We review this in the next section.
\section{$\ell^2$ Command Design by Control Theoretic Smoothing Splines}
\label{sec:l2}
For the problem given in section \ref{sec:prob}, the following $L^2$-regularized cost function
was considered in \cite{SunEgeMar00}:
\begin{equation}
 \J_2(u) := \emp(u) + \mu\reg_2(u),\quad \reg_2(u):=\int_0^T u(t)^2 \dd t.
 \label{eq:J2}
\end{equation}
The optimal control $u^\ast_2$ which minimizes $\J_2(u)$ is given
by a linear combination of the following functions called control theoretic splines \cite{SunEgeMar00,EgeMar}:
\begin{figure}[tbp]
\begin{center}
\unitlength 0.1in
\begin{picture}( 31.0000, 11.2000)(  1.3400,-11.2000)
%
{\color[named]{Black}{%
\special{pn 8}%
\special{pa 354 1120}%
\special{pa 354 0}%
\special{fp}%
\special{sh 1}%
\special{pa 354 0}%
\special{pa 334 68}%
\special{pa 354 54}%
\special{pa 374 68}%
\special{pa 354 0}%
\special{fp}%
}}%
%
{\color[named]{Black}{%
\special{pn 8}%
\special{pa 194 960}%
\special{pa 3234 960}%
\special{fp}%
\special{sh 1}%
\special{pa 3234 960}%
\special{pa 3168 940}%
\special{pa 3182 960}%
\special{pa 3168 980}%
\special{pa 3234 960}%
\special{fp}%
}}%
%
{\color[named]{Black}{%
\special{pn 13}%
\special{pa 354 960}%
\special{pa 378 942}%
\special{pa 400 922}%
\special{pa 422 902}%
\special{pa 444 882}%
\special{pa 468 862}%
\special{pa 490 842}%
\special{pa 512 820}%
\special{pa 558 776}%
\special{pa 580 752}%
\special{pa 602 726}%
\special{pa 626 700}%
\special{pa 648 674}%
\special{pa 672 644}%
\special{pa 694 614}%
\special{pa 740 552}%
\special{pa 762 520}%
\special{pa 808 460}%
\special{pa 832 432}%
\special{pa 854 406}%
\special{pa 878 382}%
\special{pa 900 362}%
\special{pa 922 344}%
\special{pa 944 332}%
\special{pa 966 324}%
\special{pa 988 320}%
\special{pa 1010 322}%
\special{pa 1032 330}%
\special{pa 1054 344}%
\special{pa 1074 360}%
\special{pa 1096 380}%
\special{pa 1118 404}%
\special{pa 1140 430}%
\special{pa 1162 458}%
\special{pa 1184 488}%
\special{pa 1206 516}%
\special{pa 1230 546}%
\special{pa 1254 576}%
\special{pa 1278 604}%
\special{pa 1304 630}%
\special{pa 1330 654}%
\special{pa 1356 676}%
\special{pa 1384 694}%
\special{pa 1412 712}%
\special{pa 1440 726}%
\special{pa 1470 740}%
\special{pa 1500 750}%
\special{pa 1530 760}%
\special{pa 1560 768}%
\special{pa 1592 774}%
\special{pa 1624 780}%
\special{pa 1656 784}%
\special{pa 1688 788}%
\special{pa 1754 792}%
\special{pa 1820 794}%
\special{pa 1952 794}%
\special{pa 1986 794}%
\special{pa 2018 794}%
\special{pa 2118 800}%
\special{pa 2182 808}%
\special{pa 2214 814}%
\special{pa 2244 820}%
\special{pa 2276 826}%
\special{pa 2338 840}%
\special{pa 2370 846}%
\special{pa 2432 860}%
\special{pa 2464 868}%
\special{pa 2526 882}%
\special{pa 2556 888}%
\special{pa 2588 896}%
\special{pa 2618 900}%
\special{pa 2650 906}%
\special{pa 2682 910}%
\special{pa 2714 916}%
\special{pa 2746 920}%
\special{pa 2776 922}%
\special{pa 2872 932}%
\special{pa 2904 934}%
\special{pa 2936 936}%
\special{pa 3032 942}%
\special{pa 3066 944}%
\special{pa 3074 944}%
\special{sp 0.070}%
}}%
%
{\color[named]{RedOrange}{%
\special{pn 20}%
\special{pa 1154 960}%
\special{pa 1132 942}%
\special{pa 1110 924}%
\special{pa 1064 886}%
\special{pa 1042 866}%
\special{pa 1018 844}%
\special{pa 996 824}%
\special{pa 974 802}%
\special{pa 950 778}%
\special{pa 928 754}%
\special{pa 906 730}%
\special{pa 882 702}%
\special{pa 860 674}%
\special{pa 838 644}%
\special{pa 816 614}%
\special{pa 770 548}%
\special{pa 748 514}%
\special{pa 724 482}%
\special{pa 702 450}%
\special{pa 680 422}%
\special{pa 656 394}%
\special{pa 634 370}%
\special{pa 612 350}%
\special{pa 588 334}%
\special{pa 566 324}%
\special{pa 544 318}%
\special{pa 520 320}%
\special{pa 498 326}%
\special{pa 476 340}%
\special{pa 454 360}%
\special{pa 430 384}%
\special{pa 408 410}%
\special{pa 386 440}%
\special{pa 362 470}%
\special{pa 354 480}%
\special{sp}%
}}%
%
{\color[named]{RedOrange}{%
\special{pn 8}%
\special{pa 1154 960}%
\special{pa 1154 472}%
\special{dt 0.045}%
\special{pa 1154 472}%
\special{pa 354 472}%
\special{dt 0.045}%
}}%
%
{\color[named]{RedOrange}{%
\special{pn 20}%
\special{pa 1150 960}%
\special{pa 3070 960}%
\special{fp}%
}}%
\put(11.5400,-10.0000){\makebox(0,0)[lt]{$t_i$}}%
\put(16.3400,-6.8000){\makebox(0,0)[lb]{$P(t)=\vc{c}^\top e^{At}\vc{b}$}}%
\put(5.2200,-2.8800){\makebox(0,0)[lb]{\textcolor{red}{$g_i(t)$}}}%
\put(3.1400,-10.0000){\makebox(0,0)[rt]{$0$}}%
\end{picture}%
\end{center}
\caption{Control theoretic spline $g_i(t)$ (solid) and the impulse response $P(t)$ of the plant $P$ (dots).}
\label{fig:base_function}
\end{figure}
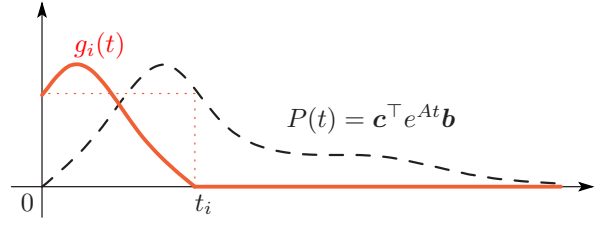
\begin{equation}
 g_i(t) := 
	\begin{cases} 
	 \vc{c}^\top \e^{A(t_i-t)}\vc{b},&\quad \text{if}~~t_i>t,\\
	 0,&\quad \text{if}~~t_i\leq t,
	\end{cases}
 \label{eq:gi}
\end{equation}
see Fig.~\ref{fig:base_function}.
More precisely, the optimal control for (\ref{eq:J2}) is given by
\begin{align}
 u_2^\ast(t) &= \sum_{i=1}^N \theta_i g_{i}(t) = \vc{g}(t)^\top\thetaQ,
 \label{eq:optimal_u2}\\
 \thetaQ &:= (\mu I + G)^{-1}\vcyref,
 \label{eq:theta2}
\end{align}
where  $\vc{g}(t) := [g_1(t),\ldots,g_N(t)]^\top$,
$\vcyref:=[\yref_1,\ldots, \yref_N]^\top$, and
$G$ is the Grammian matrix of $\{g_1,\ldots,g_N\}$,
defined by $[G]_{ij}:=\ipr{g_i}{g_j}$, $i,j=1,\ldots,N$.
\begin{figure}[tbp]
\begin{center}
\unitlength 0.1in
\begin{picture}( 30.1500,  4.8000)(  3.0000, -6.0000)
%
\special{pn 8}%
\special{pa 600 200}%
\special{pa 1600 200}%
\special{pa 1600 600}%
\special{pa 600 600}%
\special{pa 600 200}%
\special{fp}%
%
\special{pn 8}%
\special{pa 1600 400}%
\special{pa 2000 400}%
\special{dt 0.045}%
\special{sh 1}%
\special{pa 2000 400}%
\special{pa 1934 380}%
\special{pa 1948 400}%
\special{pa 1934 420}%
\special{pa 2000 400}%
\special{fp}%
%
\special{pn 8}%
\special{pa 2006 200}%
\special{pa 2406 200}%
\special{pa 2406 600}%
\special{pa 2006 600}%
\special{pa 2006 200}%
\special{fp}%
%
\special{pn 8}%
\special{pa 2406 400}%
\special{pa 2606 400}%
\special{fp}%
\special{sh 1}%
\special{pa 2606 400}%
\special{pa 2538 380}%
\special{pa 2552 400}%
\special{pa 2538 420}%
\special{pa 2606 400}%
\special{fp}%
%
\special{pn 8}%
\special{pa 2610 200}%
\special{pa 3010 200}%
\special{pa 3010 600}%
\special{pa 2610 600}%
\special{pa 2610 200}%
\special{fp}%
%
\special{pn 8}%
\special{pa 3016 400}%
\special{pa 3316 400}%
\special{fp}%
\special{sh 1}%
\special{pa 3316 400}%
\special{pa 3248 380}%
\special{pa 3262 400}%
\special{pa 3248 420}%
\special{pa 3316 400}%
\special{fp}%
\put(11.0000,-4.0000){\makebox(0,0){$\Dtheta$}}%
\put(3.0000,-3.5000){\makebox(0,0)[lb]{$\vcyref$}}%
\put(17.0000,-3.5000){\makebox(0,0)[lb]{$\thetaQ$}}%
\put(24.4000,-3.5000){\makebox(0,0)[lb]{$u_2^\ast$}}%
\put(32.6500,-3.5000){\makebox(0,0)[lb]{$y$}}%
\put(22.0500,-4.0000){\makebox(0,0){$\vc{g}(t)$}}%
\put(28.1500,-4.0000){\makebox(0,0){$P$}}%
%
\special{pn 8}%
\special{pa 300 400}%
\special{pa 600 400}%
\special{fp}%
\special{sh 1}%
\special{pa 600 400}%
\special{pa 534 380}%
\special{pa 548 400}%
\special{pa 534 420}%
\special{pa 600 400}%
\special{fp}%
\end{picture}%
\end{center}
\caption{Remote-controlled system optimized with $\J_2(u)$ in (\ref{eq:J2}).
The vector $\thetaQ$ 
is transmitted through a communication channel.}
\label{fig:rc_system}
\end{figure}
\section{$\elll$ Command Design for Sparse Remote Control}
In remote-controlled systems,
we transmit the control input $u=\{u(t)\}_{t\in[0,T]}$ to the system $P$
through a communication channel.
Since $\{u(t)\}_{t\in[0,T]}$ is a continuous-time signal,
we should discretize it.

An easy way to communicate information on the input signal is
to transmit the data $\vcyref$ itself,
and produce the input $u(t)$ by the formulae (\ref{eq:optimal_u2})
and (\ref{eq:theta2}) at the receiver side.
The vector $\vcyref$ is just an $N$-dimensional one, and much easier
to transmit than the infinite-dimensional vector $\{u(t)\}_{t\in[0,T]}$.

An alternative method consists in transmitting the coefficient vector $\thetaQ$
given in (\ref{eq:theta2})
instead of the continuous-time signal $u$.
This procedure is shown in Fig.~\ref{fig:rc_system}.
In this procedure, we fix the sampling instants $t_1,\ldots,t_N$ and
the vector $\vcyref$ is given.
We first compute the parameter vector $\thetaQ$ by (\ref{eq:theta2}),
and transmit this through a communication channel.
The transmitted vector is received at the receiver,
and then the control signal $u^\ast_2(t)$ is computed by (\ref{eq:optimal_u2}),
and applied to the plant $P$.
We assume that the time instants $t_1,\ldots,t_N$ are
shared at the transmitter and the receiver.

A problem of the above-mentioned strategies is that the communication channel
is band-limited and therefore the vector to be transmitted
has to be first quantized and encoded.
To solve this, we will seek a {\it sparse representation} of the transmitted vector $\vc{\theta}$
in accordance with the notion of compressed sensing
\cite{CanWak08,Ela}.

Define a subspace $\V$ of $L^2[0,T]$ by
\begin{equation}
 \V := \left\{u\in L^2[0,T]: u = \sum_{j=1}^M \theta_j \phi_j,~ \theta_i \in \re\right\},
 \label{eq:VM}
\end{equation}
where $\phi_1, \ldots, \phi_M$ are linearly independent vectors in $L^2[0,T]$.
Note that if $M=N$ and $\phi_i=g_i$, $i=1,\ldots,N$ defined in (\ref{eq:gi}),
the optimal control $u^\ast_2(t)$ in (\ref{eq:optimal_u2}) belongs to this subspace%
\footnote{The functions $\{g_1,\ldots,g_N\}$ are linearly independent \cite{SunEgeMar00}.}.
We assume that the control $u$ is in $\V$, that is, we find a control $u$
in this subset.
Under this assumption, the squared-error cost function $\emp(u)$ is represented by
\begin{equation}
  \emp(u)=\sum_{i=1}^N (y(t_i)-\yref_i)^2
    = \left\|\Phi\vc{\theta} - \vcyref\right\|_2^2,
  \label{eq:l2error}
\end{equation}
where $[\Phi]_{ij}=\ipr{g_i}{\phi_j}$,
$i=1,\ldots,N$, $j=1,\ldots,M$.
To induce sparsity in $\vc{\theta}$, we adopt $\ell^1$ penalty on $\vc{\theta}$
and introduce the following mixed $\elll$ cost function:
\begin{equation}
\J_1(\vc{\theta}) := \frac{1}{2}\left\|\Phi\vc{\theta} - \vcyref\right\|_2^2 + \kappa \|\vc{\theta}\|_1.
 \label{eq:J1}
\end{equation}
Note that if $\|\phi_j\|_1=1$ for $j=1,\ldots,M$,
then the cost function (\ref{eq:J1}) is an upper bound of the following
$L^1$-$L^2$ cost function:
\[
 \J_1(u) = \frac{1}{2}\emp(u) + \kappa \reg_1(u),\quad \reg_1(u) = \int_0^T |u(t)|\dd t.
\]

As mentioned in the introduction,
the $\ell^1$-regularized least-squares optimization is a good approximation
to one regularized by the $\ell^0$ norm which counts the nonzero elements in $\vc{\theta}$.
Although the solution which minimizes $\J_1(\vc{\theta})$ cannot be represented
analytically as in (\ref{eq:optimal_u2}),
we can compute an approximated solution by using a fast numerical algorithm.
The algorithm is described in the next section.
By using this solution, say $\thetaS$,
the optimal control $u_1^\ast$ can be obtained from
\[
 u_1^\ast(t) = \sum_{i=1}^N \theta^\ast_i \phi_i(t) 
  = \vc{\phi}(t)^\top \thetaS,\quad t\in[0,T].
\]
\section{Sparse Representation by $\elll$ Optimization}
\label{sec:sparse}
We here describe a fast algorithm for obtaining the optimal vector $\thetaS$.
First, we consider a general case of optimization.
Next, we simplify the design procedure in a special case.
\subsection{General case}
\label{subsec:sparse-general}
The cost function (\ref{eq:J1}) is convex in $\theta$ and hence
the optimal value $\thetaS$ uniquely exists.
However, an analytical expression as in (\ref{eq:theta2})
for this optimal vector
is unknown except when the matrix $\Phi$ is unitary.
To obtain the optimal vector $\thetaS$,
one can use an iteration method.
Recently, a very fast algorithm for the optimal $\elll$ solution
has been proposed, which is called {\it iterative shrinkage} \cite{BecTeb09,ZibEla10}.

\begin{figure}[tb]
\begin{center}
\unitlength 0.1in
\begin{picture}( 24.8000, 20.4000)( -0.8000,-20.0000)
%
\special{pn 8}%
\special{pa 400 1000}%
\special{pa 2400 1000}%
\special{fp}%
\special{sh 1}%
\special{pa 2400 1000}%
\special{pa 2334 980}%
\special{pa 2348 1000}%
\special{pa 2334 1020}%
\special{pa 2400 1000}%
\special{fp}%
%
\special{pn 8}%
\special{pa 1400 2000}%
\special{pa 1400 0}%
\special{fp}%
\special{sh 1}%
\special{pa 1400 0}%
\special{pa 1380 68}%
\special{pa 1400 54}%
\special{pa 1420 68}%
\special{pa 1400 0}%
\special{fp}%
%
\special{pn 20}%
\special{pa 400 1600}%
\special{pa 1000 1000}%
\special{fp}%
\special{pa 1000 1000}%
\special{pa 1800 1000}%
\special{fp}%
\special{pa 1800 1000}%
\special{pa 2400 400}%
\special{fp}%
\put(14.5000,-10.5000){\makebox(0,0)[lt]{$0$}}%
\put(18.5000,-10.5000){\makebox(0,0)[lt]{$\kappa/c$}}%
\put(10.0000,-9.5000){\makebox(0,0)[rb]{$-\kappa/c$}}%
\put(24.0000,-9.5000){\makebox(0,0)[lb]{$\theta$}}%
\end{picture}%
\end{center}
\caption{Nonlinear function $\sgn(\theta)(|\theta|-\kappa/c)_+$}
\label{fig:nonlinear}
\end{figure}
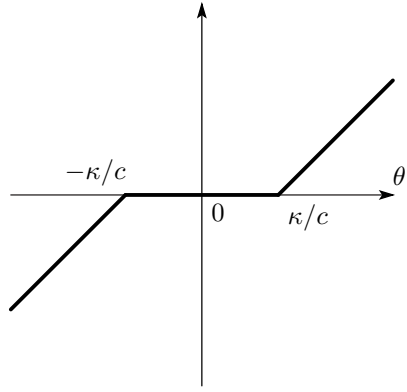

This algorithm is given by the following:
Give an initial value $\vc{\theta}[0]\in\re^M$, and let $\beta[1]=1$, $\vc{\theta}'[1]=\vc{\theta}[0]$.
Fix a constant $c$ such that $c>\|\Phi\|^2:=\lambda_{\max}(\Phi^\top\Phi)$.
Execute the following iteration%
\footnote{Several methods have been proposed
for the iterative shrinkage \cite{ZibEla10}.
The algorithm given here is called
FISTA (Fast Iterative Shrinkage-Thresholding Algorithm) \cite{BecTeb09}.}:
\begin{equation}
 \begin{split}
  \vc{\theta}[j] &= \Sr_{\kappa/c}\left(\frac{1}{c}\Phi^\top(\vcyref-\Phi\vc{\theta}'[j])+\vc{\theta}'[j]\right),\\
  \beta[j+1] &= \frac{1+\sqrt{1+4\beta[j]^2}}{2},\\
  \vc{\theta}'[j+1] &= \vc{\theta}[j] + \frac{\beta[j]-1}{\beta[j+1]}(\vc{\theta}[j] - \vc{\theta}[j-1]),\\
  j &= 1,2,\ldots,
 \end{split}
\label{eq:fista}
\end{equation}
where the function $\Sr_{\kappa/c}$ is defined for $\vc{\theta}=[\theta_1,\ldots,\theta_M]^\top$ by
\[
  \Sr_{\kappa/c}(\vc{\theta}) := 
	\begin{bmatrix}
	 \sgn(\theta_1)(|\theta_1|-\kappa/c)_+\\
	 \vdots\\
	 \sgn(\theta_M)(|\theta_M|-\kappa/c)_+
	\end{bmatrix}.
\]
The nonlinear function $\sgn(\theta)(|\theta|-\kappa/c)$ in $\Sr_{\kappa/c}$
is shown in Fig.~\ref{fig:nonlinear}.
If $c>\|\Phi\|^2$, the above algorithm converges to
the optimal solution minimizing the $\elll$ cost function
(\ref{eq:J1}) for any initial value $\vc{\theta}[0]\in\re^{M}$
with a worst-case convergence rate $O(1/j^2)$ \cite{DauDefMol04,BecTeb09}.
The above algorithm is very simple and fast;
it can be effectively implemented in digital devices,
which leads to a real-time computation of a sparse vector $\thetaS$.
\begin{figure}[tbp]
\begin{center}
\unitlength 0.1in
\begin{picture}( 30.1500,  4.8000)(  3.0000, -6.0000)
%
\special{pn 8}%
\special{pa 600 200}%
\special{pa 1600 200}%
\special{pa 1600 600}%
\special{pa 600 600}%
\special{pa 600 200}%
\special{fp}%
%
\special{pn 8}%
\special{pa 1600 400}%
\special{pa 2000 400}%
\special{dt 0.045}%
\special{sh 1}%
\special{pa 2000 400}%
\special{pa 1934 380}%
\special{pa 1948 400}%
\special{pa 1934 420}%
\special{pa 2000 400}%
\special{fp}%
%
\special{pn 8}%
\special{pa 2006 200}%
\special{pa 2406 200}%
\special{pa 2406 600}%
\special{pa 2006 600}%
\special{pa 2006 200}%
\special{fp}%
%
\special{pn 8}%
\special{pa 2406 400}%
\special{pa 2606 400}%
\special{fp}%
\special{sh 1}%
\special{pa 2606 400}%
\special{pa 2538 380}%
\special{pa 2552 400}%
\special{pa 2538 420}%
\special{pa 2606 400}%
\special{fp}%
%
\special{pn 8}%
\special{pa 2610 200}%
\special{pa 3010 200}%
\special{pa 3010 600}%
\special{pa 2610 600}%
\special{pa 2610 200}%
\special{fp}%
%
\special{pn 8}%
\special{pa 3016 400}%
\special{pa 3316 400}%
\special{fp}%
\special{sh 1}%
\special{pa 3316 400}%
\special{pa 3248 380}%
\special{pa 3262 400}%
\special{pa 3248 420}%
\special{pa 3316 400}%
\special{fp}%
\put(11.0000,-4.0000){\makebox(0,0){FISTA}}%
\put(3.0000,-3.5000){\makebox(0,0)[lb]{$\vcyref$}}%
\put(16.4000,-3.5000){\makebox(0,0)[lb]{$\thetaS$}}%
\put(24.4000,-3.5000){\makebox(0,0)[lb]{$u_1^\ast$}}%
\put(32.6500,-3.5000){\makebox(0,0)[lb]{$y$}}%
\put(22.0500,-4.0000){\makebox(0,0){$\vc{g}(t)$}}%
\put(28.1500,-4.0000){\makebox(0,0){$P$}}%
%
\special{pn 8}%
\special{pa 300 400}%
\special{pa 600 400}%
\special{fp}%
\special{sh 1}%
\special{pa 600 400}%
\special{pa 534 380}%
\special{pa 548 400}%
\special{pa 534 420}%
\special{pa 600 400}%
\special{fp}%
\end{picture}%
\end{center}
\caption{Remote-controlled system optimized with $\J_1(\vc{\theta})$ in (\ref{eq:J1}).
The vector $\thetaS$ minimizing (\ref{eq:J1}) is computed by
the FISTA (Fast Iterative Shrinkage-Thresholding Algorithm) given in (\ref{eq:fista}),
and transmitted through a communication channel.}
\label{fig:rc_system_fista}
\end{figure}
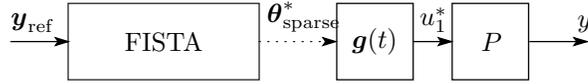
\subsection{The case $\Phi=G$}
\label{subsec:special-case}
We here assume $M=N$ and $\phi_i=g_i$, $i=1,2,\ldots,N$,
that is, $\Phi=G$.
Since $g_1, \ldots, g_N$ are linearly independent vectors in $L^2[0,T]$,
the Grammian matrix $\Phi=G$ is non-singular.
Let the control input $u$ be
\[
 u(t) = \sum_{i=1}^N \theta_i g_i(t) = \vc{g}(t)^\top \vc{\theta},
\]
and let $\vc{\eta}:=\Phi\vc{\theta}$.
Then, by (\ref{eq:l2error}) we have
\[
 \sum_{i=1}^N \left(y(t_i)-\yref_i\right)^2 = \|\vc{\eta}-\vcyref\|_2^2.
\]
Consider the following $\elll$ cost function:
\begin{equation}
 J(\vc{\eta}) = \nu\|\vc{\eta}\|_1 + \frac{1}{2}\|\vc{\eta}-\vcyref\|_2^2.
 \label{eq:Jzeta}
\end{equation}
The optimal solution $\etaS$
minimizing this cost function is given analytically by
\begin{equation}
 \etaS = \Sr_\nu(\vcyref).\label{eq:etaS}
\end{equation}
Then we transmit this optimal vector $\etaS$,
and at the receiver we reconstruct the optimal control by
$u_1^\ast(t) = \vc{g}(t)^\top\Phi^{-1}\etaS$.
Fig.~\ref{fig:rc_system_1} shows the remote-controlled system with the optimizer $\etaS$.
\begin{figure}[tbp]
\begin{center}
\unitlength 0.1in
\begin{picture}( 30.0000,  4.2000)(  5.0000, -6.0000)
%
\special{pn 8}%
\special{pa 500 400}%
\special{pa 800 400}%
\special{fp}%
\special{sh 1}%
\special{pa 800 400}%
\special{pa 734 380}%
\special{pa 748 400}%
\special{pa 734 420}%
\special{pa 800 400}%
\special{fp}%
%
\special{pn 8}%
\special{pa 800 200}%
\special{pa 1200 200}%
\special{pa 1200 600}%
\special{pa 800 600}%
\special{pa 800 200}%
\special{fp}%
%
\special{pn 8}%
\special{pa 1600 200}%
\special{pa 2000 200}%
\special{pa 2000 600}%
\special{pa 1600 600}%
\special{pa 1600 200}%
\special{fp}%
%
\special{pn 8}%
\special{pa 2000 400}%
\special{pa 2200 400}%
\special{fp}%
\special{sh 1}%
\special{pa 2200 400}%
\special{pa 2134 380}%
\special{pa 2148 400}%
\special{pa 2134 420}%
\special{pa 2200 400}%
\special{fp}%
%
\special{pn 8}%
\special{pa 2200 200}%
\special{pa 2600 200}%
\special{pa 2600 600}%
\special{pa 2200 600}%
\special{pa 2200 200}%
\special{fp}%
%
\special{pn 8}%
\special{pa 2600 400}%
\special{pa 2800 400}%
\special{fp}%
\special{sh 1}%
\special{pa 2800 400}%
\special{pa 2734 380}%
\special{pa 2748 400}%
\special{pa 2734 420}%
\special{pa 2800 400}%
\special{fp}%
%
\special{pn 8}%
\special{pa 2800 200}%
\special{pa 3200 200}%
\special{pa 3200 600}%
\special{pa 2800 600}%
\special{pa 2800 200}%
\special{fp}%
%
\special{pn 8}%
\special{pa 3200 400}%
\special{pa 3500 400}%
\special{fp}%
\special{sh 1}%
\special{pa 3500 400}%
\special{pa 3434 380}%
\special{pa 3448 400}%
\special{pa 3434 420}%
\special{pa 3500 400}%
\special{fp}%
\put(30.0000,-4.0000){\makebox(0,0){$P$}}%
\put(24.0000,-4.0000){\makebox(0,0){$\vc{g}(t)$}}%
\put(18.0000,-4.0000){\makebox(0,0){$\Phi^{-1}$}}%
\put(10.0000,-4.0000){\makebox(0,0){$\Sr_\nu$}}%
\put(5.0000,-3.5000){\makebox(0,0)[lb]{$\vcyref$}}%
\put(12.3000,-3.5000){\makebox(0,0)[lb]{$\etaS$}}%
\put(26.2500,-3.5000){\makebox(0,0)[lb]{$u_1^\ast$}}%
\put(33.5000,-3.5000){\makebox(0,0)[lb]{$y$}}%
%
\special{pn 8}%
\special{pa 1200 400}%
\special{pa 1600 400}%
\special{dt 0.045}%
\special{sh 1}%
\special{pa 1600 400}%
\special{pa 1534 380}%
\special{pa 1548 400}%
\special{pa 1534 420}%
\special{pa 1600 400}%
\special{fp}%
\end{picture}%
\end{center}
\caption{Remote-controlled system optimized with $J(\vc{\eta})$ in (\ref{eq:Jzeta}).
The vector $\vc{\eta}$ is transmitted through a communication channel.}
\label{fig:rc_system_1}
\end{figure}
In this case, we compute (\ref{eq:etaS}) only one time,
while in the general case considered in Section \ref{subsec:sparse-general}
we should execute the iteration algorithm (\ref{eq:fista}).

\section{Example}
We here show an example of the sparse command generator.
The state-space matrices of the controlled plant $P$ is assumed to be
\[
 A = \begin{bmatrix}0&1\\-1&-2\end{bmatrix},\quad
 B = \begin{bmatrix}0\\1\end{bmatrix},\quad
 C = \begin{bmatrix}1&0\end{bmatrix}.
\]
Note that the transfer function of the plant $P$ is $1/(s+1)^2$.
The sampling instants are given by
$t_i = i\times \pi/6$, $i=1,2,\ldots,12$,
and the data $Y_1,\ldots,Y_{12}$ is given by $Y_i = \sin t_i$,
that is, we try to track the sine function $y(t)=\sin t$
in one period $[0,2\pi]$.
We assume the base functions $\phi_i$
in the subspace $\V$ in (\ref{eq:VM}) are the same as $g_i$'s,
that is, we consider the case $\Phi=G$ discussed in Section \ref{subsec:special-case}.
We design three signals to be transmitted:
the $\ell^2$-optimized vector $\thetaQ$ in (\ref{eq:theta2}),
the sparse vector $\thetaS$ given in subsection \ref{subsec:sparse-general},
and the sparse vector $\etaS$ in (\ref{eq:etaS}).
We set the regularization parameters $\mu=0.01$,
$\kappa=0.001$, and $\nu=0.05$,
see equations (\ref{eq:J2}), (\ref{eq:J1}) and (\ref{eq:Jzeta}).

The obtained vectors are shown in Table \ref{table:vectors}.
\begin{table}{tbp}
\caption{Designed vectors}
\label{table:vectors}
\begin{center}
\begin{tabular}{|c|c|c|c|}\hline
$\thetaQ$ & $\thetaS$ & $\etaS$ & $\vcyref$\\\hline
    9.7994&    9.6727&    0.4500&    0.5000\\
    2.7995&    4.5626&    0.8160&    0.8660\\
    1.6544&         0&    0.9500&    1.0000\\
    1.6695&    2.9973&    0.8160&    0.8660\\
    1.0358&         0&    0.4500&    0.5000\\
    0.0059&         0&         0&    0.0000\\
   -1.0231&         0&   -0.4500&   -0.5000\\
   -1.7456&   -2.8678&   -0.8160&   -0.8660\\
   -2.0234&   -0.6316&   -0.9500&   -1.0000\\
   -2.2424&   -4.8575&   -0.8160&   -0.8660\\
   -2.4153&         0&   -0.4500&   -0.5000\\
    5.1813&    4.4185&         0&   -0.0000\\\hline
\end{tabular}
\end{center}
\end{table}
We can see that the vector $\thetaS$ is the sparsest due to the sparsity-inducing
approach. The second sparsest vector is $\etaS$ which converts small elements in $\vcyref$
to 0. The vector $\thetaQ$ is not sparse.

\begin{figure}[tbp]
\begin{center}
\includegraphics[width=0.9\linewidth]{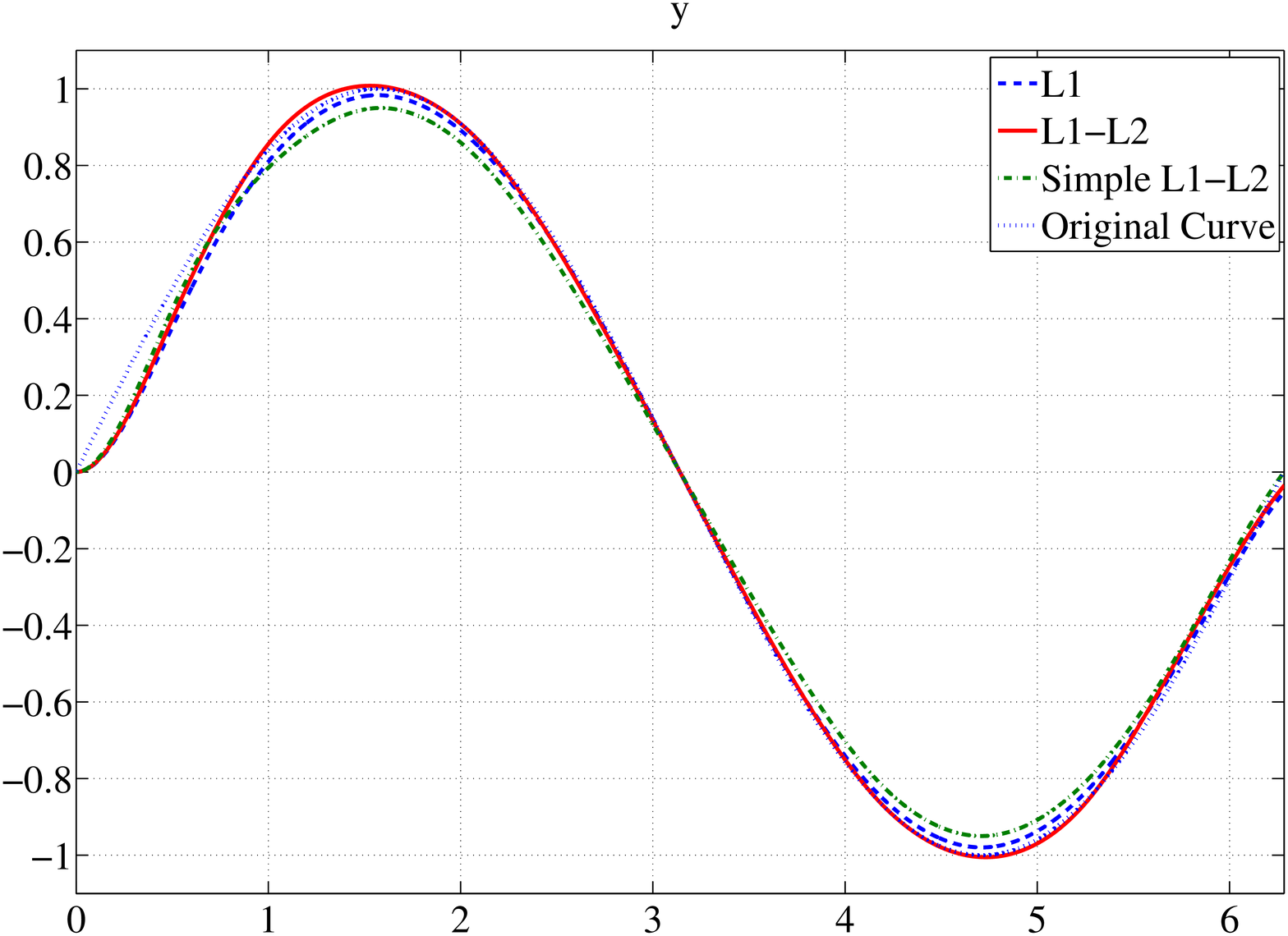}
\end{center}
\caption{The original curve (dots) and outputs:
by $\ell^2$-optimal $\thetaQ$ (dash), $\elll$-optimal $\thetaS$ (solid),
and simple $\elll$-optimal $\etaS$ (dash-dots).}
\label{fig:y}
\end{figure}

Fig.~\ref{fig:y} shows the plant outputs obtained by the above vectors.
\begin{figure}[tbp]
\begin{center}
\includegraphics[width=0.9\linewidth]{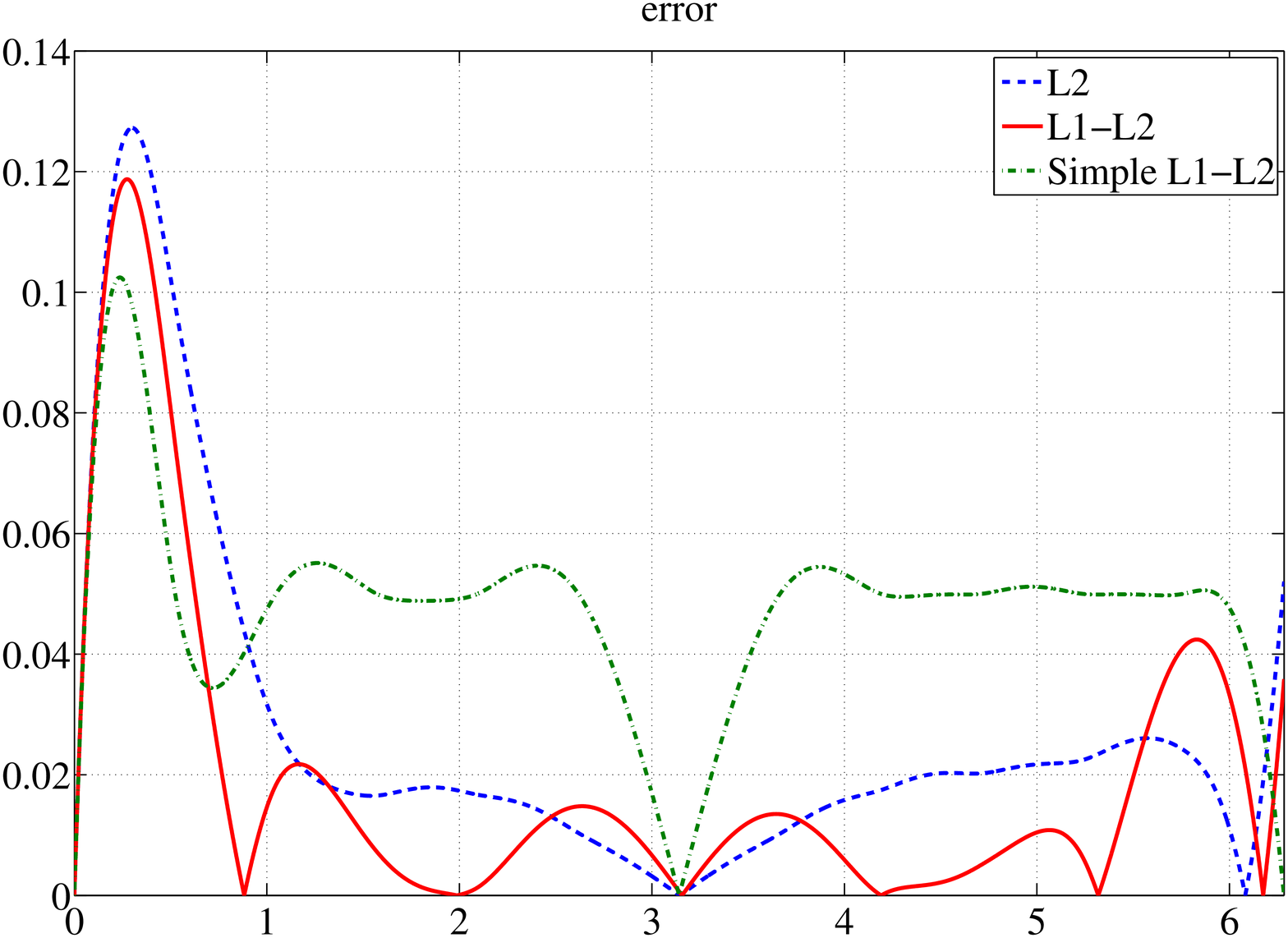}
\end{center}
\caption{The reconstruction errors:
by $\ell^2$-optimal $\thetaQ$ (dash), $\elll$-optimal $\thetaS$ (solid),
and simple $\elll$-optimal $\etaS$ (dash-dots).}
\label{fig:error}
\end{figure}
The transient responses show relatively large errors because of the phase delay in the plant $P(s)=1/(s+1)^2$.
Despite of sparsity in $\thetaS$ and $\etaS$,
the performances of the reconstructed signals are comparable to 
that of the $\ell^2$-optimal reconstruction by $\thetaQ$.
To see the difference between these performances more precisely,
we draw the reconstruction errors in Fig.~\ref{fig:error}.
We can see that the errors by $\thetaQ$ and $\thetaS$ are almost comparable,
and the error by $\etaS$ is relatively large.

\begin{figure}[tbp]
\begin{center}
\includegraphics[width=0.9\linewidth]{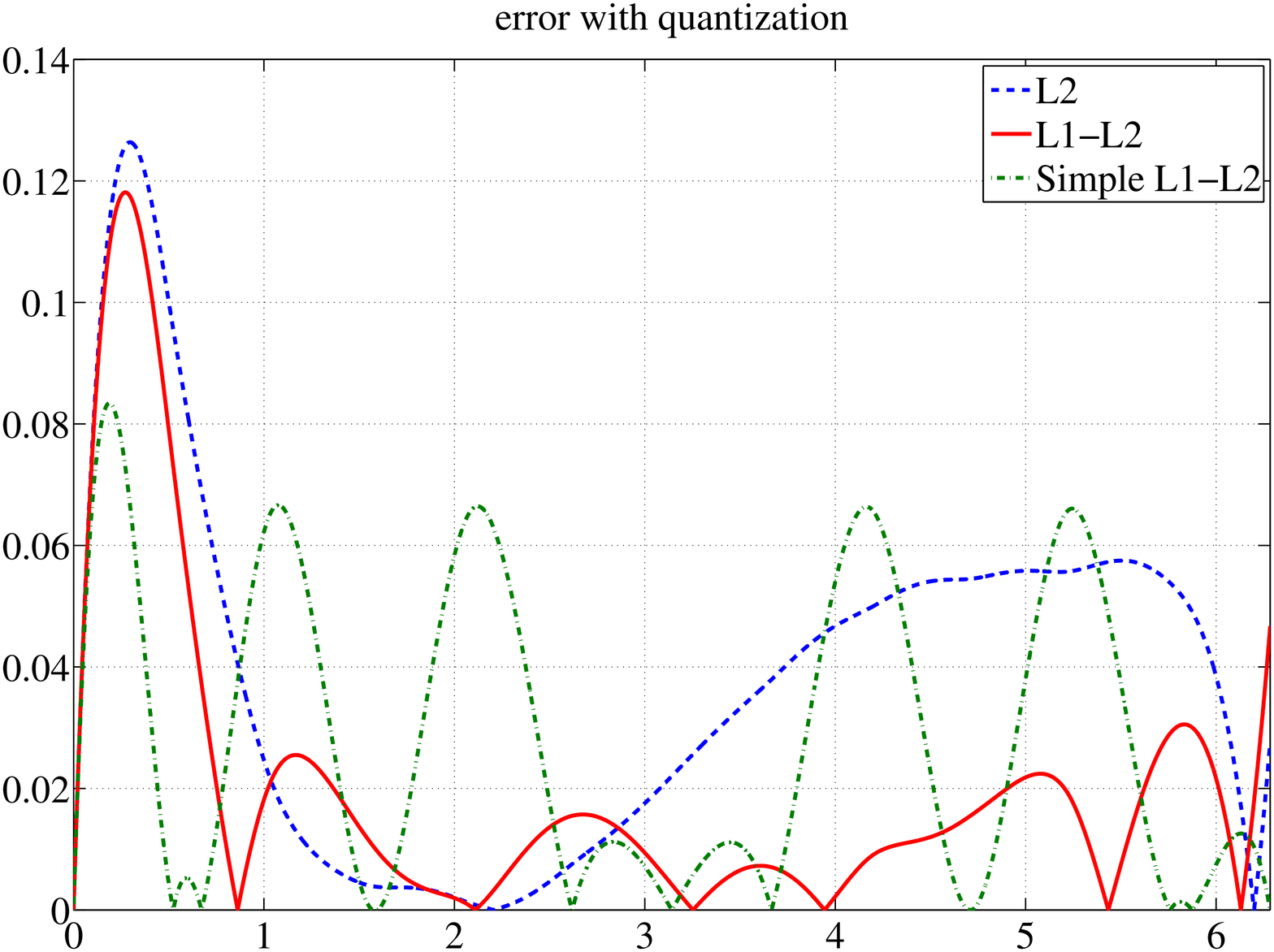}
\end{center}
\caption{The reconstruction errors with quantization:
by $\ell^2$-optimal $\hat{\vc{\theta}}_2^\ast$ (dash), $\elll$-optimal $\hat{\vc{\theta}}_{\text{sparse}}^\ast$ (solid),
and simple $\elll$-optimal $\hat{\vc{\eta}}_{\text{sparse}}^\ast$ (dash-dots).}
\label{fig:errorQ}
\end{figure}

Then we consider quantization.
We use the uniform quantizer with step size $0.1$ and simulate the output reconstruction.
Table \ref{table:Qvectors} shows the quantized vectors.
Fig.~\ref{fig:errorQ} shows the reconstruction error under quantization.
The errors by the sparse vectors $\thetaS$ and $\etaS$ still remains small
while the $\ell^2$-optimal reconstruction shows errors affected by quantization.
This is because the zero-valued elements in the sparse vectors do not suffer from
any quantization distortion.
\begin{table}{tbp}
\caption{Quantized vectors}
\label{table:Qvectors}
\begin{center}
\begin{tabular}{|c|c|c|c|}\hline
$Q(\thetaQ)$ & $Q(\thetaS)$ & $Q(\etaS)$ & $Q(\vcyref)$\\\hline
    9.8&    9.7&    0.5&    0.5\\
    2.8&    4.6&    0.8&    0.9\\
    1.7&    0.0&    1.0&    1.0\\
    1.7&    3.0&    0.8&    0.9\\
    1.0&    0.0&    0.5&    0.5\\
    0.0&    0.0&    0.0&    0.0\\
   -1.0&    0.0&   -0.5&   -0.5\\
   -1.7&   -2.9&   -0.8&   -0.9\\
   -2.0&   -0.6&   -1.0&   -1.0\\
   -2.2&   -4.9&   -0.8&   -0.9\\
   -2.4&    0.0&   -0.5&   -0.5\\
    5.2&    4.4&    0.0&    0.0\\
\hline
\end{tabular}
\end{center}
\end{table}
\section{Conclusion}
In this paper, we have proposed to use sparse representation for command generation
in remote control by $\elll$ optimization.
An example illustrates the effectiveness of the proposed method.
Future work may include the study of advantages of sparse representation
in view of information theory.


\end{document}